\documentclass[twocolumn,showpacs,preprintnumbers,amsmath,amssymb]{revtex4}
\usepackage{dcolumn}
\usepackage{bm}
\usepackage{graphicx}
\usepackage{amsfonts}
\usepackage{amsmath}
\newcommand{\bfb}{\cal}
\newcommand{\Char}{\mathop{\rm Char}\nolimits}
\newcommand{\WF}{\mathop{\rm WF}\nolimits}
\newcommand{\Spec}{\mathop{\rm Spec}\nolimits}

\begin{document}
\title{Algebra of distributions of quantum-field densities and space-time properties }
\author{Leonid Lutsev}
\email{l_lutsev@mail.ru}

\affiliation{Ioffe Physical-Technical Institute,
Russian Academy of Sciences, 194021, St. Petersburg, Russia }
\date{\today}

\begin{abstract}
In this paper we consider properties of the space-time manifold
${\bfb M}$ caused by the proposition that, according to the scheme
theory, the manifold ${\bfb M}$ is locally isomorphic to the
spectrum of the algebra ${\bfb A}$, ${\bfb M}\cong\Spec({\bfb A})$,
where ${\bfb A}$ is the commutative algebra of distributions of
quantum-field densities. In order to determine the algebra ${\bfb
A}$, it is necessary to define multiplication on densities and to
eliminate those densities, which cannot be multiplied. This leads to
essential restrictions imposed on densities and on space-time
properties. It is found that the only possible case, when the
commutative algebra ${\bfb A}$ exists, is the case, when the quantum
fields are in the space-time manifold ${\bfb M}$ with the structure
group $SO(3,1)$ (Lorentz group). The algebra ${\bfb A}$ consists of
distributions of densities with singularities in the closed future
light cone subset. On account of the local isomorphism ${\bfb
M}\cong\Spec({\bfb A})$, the quantum fields exist only in the
space-time manifold with the one-dimensional arrow of time. In the
fermion sector the restrictions caused by the possibility to define
the multiplication on the densities of spinor fields can explain the
chirality violation. It is found that for bosons in the Higgs sector
the charge conjugation symmetry violation on the densities of states
can be observed. This symmetry violation can explain the
matter-antimatter imbalance.
\end{abstract}



\maketitle

\section{INTRODUCTION}

The origin of the arrow of time, the possibility of physics in
multiple time dimensions, the violation of the parity principle, and
the matter-antimatter imbalance are ones of the most exciting and
difficult challenges of physics.

Physics in multiple time dimensions leads to new insights and, at
the same time, contains theoretical problems. Extra time dimensions
give new hidden symmetries that conventional one time physics does
not capture, implying the existence of a more unified formulation of
physics that naturally supplies the hidden
information~\cite{Bars2000,Bars06,Bars10}. At the same time, it
notes that all but the $(3+1)$-dimensional one might correspond to
"dead worlds", devoid of observers, and we should find ourselves
inhabiting a $(3+1)$-dimensional space-time~\cite{Teg97}. The
natural description of the $(3+1)$-space-time with the
one-dimensional time can be provided on the base of the Clifford
geometric algebra~\cite{Chap16}. In the opposite case of
multidimensional time, the violation of the causal structure of the
space-time and the movement backwards in the time dimensions are
possible~\cite{Velev12}. A particle can move in the causal region
faster than the speed of light in vacuum. This leads to
contradictoriness of the multidimensional time theory and, at
present, these problems have not been solved.

The arrow of time is the one-way property of time which has no
analogue in space. The asymmetry of time is explained by large
numbers of theoretical models -- by the Second law of thermodynamics
(the thermodynamic arrow of time), by the direction of the universe
expansion (the cosmological arrow), by the quantum uncertainty and
entanglement of quantum states (the quantum source of time), and by
the perception of a continuous movement from the known (past) to the
unknown (future) (the psychological time
arrow)~\cite{Leb93,Kief95,Zeh10,Jen10,Mers12,Bous12,Smol13}. At
present, there is not a satisfactory explanation of the arrow of
time and this problem is far from being solved.

The discrete symmetry of the space reflection $P$ of the space-time
and the charge conjugation $C$ may be used to characterize the
properties of chiral systems. The violation of the space reflection
exhibits as the chiral symmetry breaking -- only left-handed
particles and right-handed anti-particles could be
observed~\cite{Huang82,Cheng84,Barron12}. The matter-antimatter
imbalance remains as one of the unsolved problems. The amount of
$CP$ violation in the Standard Model is insufficient to account for
the observed baryon asymmetry of the universe. At present, the hope
to explain the matter-antimatter imbalance is set on the $CP$
violation in the Higgs
sector~\cite{Shu13,Cir16,Car16,Engl17,Good17}.

In this paper we consider the above-mentioned space-time properties
(the arrow of time, multiple time dimensions, and the chirality
violation), the violation of the charge conjugation, and find that
in the framework of the scheme theory these properties are
determined by the commutative algebra ${\bfb A}$ of quantum-field
densities. Schemes were introduced by Alexander Grothendieck with
the aim of developing the formalism needed to solve deep problems of
algebraic geometry~\cite{EGA}. This led to the evolution of the
concept of space~\cite{Cart01}. The space is associated with a
spectrum of a commutative algebra. In the case of the classical
physics, the commutative algebra is the commutative ring of
functions. In contrast with the classical physics, quantum fields
are determined by equations on
functionals~\cite{Schw51,Vas98,Lut07,Lut09}. Quantum-field densities
are linear functionals of auxiliary fields and, consequently, are
distributions. There are many restrictions to construct the
commutative algebra of distributions. In the common case,
multiplication on distributions cannot be defined and depends on
their wavefront sets. In microlocal analysis the wavefront set
$\WF(u)$ characterizes the singularities of a distribution $u$, not
only in space, but also with respect to its Fourier transform at
each point. The term wavefront was coined by Lars
H\"{o}rmander~\cite{Hor71}. It needs to note that the microlocal
analysis has resulted in the recent progress in the renormalized
quantum field theory in curved
space-time~\cite{Radz96,Brun00,Stroh02,Broud14}. In our case, the
possibility to define multiplication on distributions leads to
essential restrictions imposed on densities forming the algebra
${\bfb A}$. The spectrum of the algebra ${\bfb A}$ is locally
isomorphic to the space-time manifold ${\bfb M}$, ${\bfb
M}\cong\Spec({\bfb A})$, and characterizes its properties such as
the one-dimensional arrow of time, the chirality violation and the
structure group of the space-time manifold ${\bfb M}$. One can say
that the space-time is determined by matter.

The paper is organized as follows. In Section II we derive
differential equations for the densities of quantum fields from the
Schwinger equation and find that, in the common case, the quantum field
densities are distributions. The quantum fields contain fermion, boson (Higgs),
and gauge field components. Multiplication on the quantum-field
densities and the commutative algebra ${\bfb A}$ of distributions of
densities are considered in Section III. It is found that the only
possible case, when the commutative algebra ${\bfb A}$ of
distributions of quantum-field densities exists, is the case, when
the quantum fields are in the space-time manifold ${\bfb M}$ with
the structure group $SO(3,1)$ (Lorentz group) and the time is
one-dimensional. The asymmetry of time, the chirality violation of
spinor fields, and the charge conjugation symmetry violation in the
boson sector are the necessary conditions for the existence of the
algebra ${\bfb A}$. The quantum fields exist only in the space-time
manifold with the one-dimensional arrow of time and with chirality
and charge conjugation symmetry violations. Ideals, localization,
the spectrum of the density distribution algebra ${\bfb A}$ and the
scheme $({\bfb M},{\bfb A}_{{\bfb M}})$ are considered in Section IV.
If the algebra ${\bfb A}$ can be determined and is the component of
the scheme, then the space-time manifold ${\bfb M}$ with quantum
fields exists. Otherwise, the space-time manifold is devoid of matter
and, consequently, does not exist.

\section{QUANTUM-FIELD EQUATIONS}

Quantum fields are determined by equations on functionals. In this
section we consider singularities of the linear
components $W^{\zeta}(x)$ (densities) of the functional solution

$$G(Q)=\sum_{\zeta}\int W^{\zeta}(x)Q^{\zeta}(x)\,dx$$
$$+ \sum_{n>1}\sum_{\zeta_1\ldots\zeta_n} \int\!\!\ldots\!\!\int
W^{\zeta_1\ldots\zeta_n}(x_1\ldots x_n)Q^{\zeta_1}(x_1)$$
$$\ldots Q^{\zeta_n}(x_n)\,dx_1\ldots dx_n, $$

\noindent where $Q^{\zeta}(x)$ is the auxiliary fields. For this
purpose, we derive differential equations for the densities of
quantum fields from the Schwinger equation. Let us consider fields
$\Psi$ on the space-time manifold ${\bfb M}$

\begin{equation}
\Psi(x)=\{\Psi^{\zeta}(x)\}=\{\psi^{\alpha}(x),
\bar\psi^{\alpha}(x), \varphi^n(x), \varphi^{{+}n}(x),
A^{a}_{\mu}(x)\}, \label{eq1}
\end{equation}

\noindent where $\psi^{\alpha}(x)$, $\bar\psi^{\alpha}(x)$ are the
fermion (spinor) fields, $\varphi^n(x)$, $\varphi^{{+}n}(x)$ are the
bosons (for example, Higgs bosons), and $A^{a}_{\mu}(x)$ are the
gauge field potentials. In relation (\ref{eq1}) and
in the all following relations Greek letter indices
$\alpha$ and $\beta$ enumerate types of fermions, $\mu$, $\nu$ and $\rho$
are indices of the space-time variables, Latin letters $n$, $m$, $l$
enumerate types of bosons, and $a$, $b$, $c$ are the gauge
indices, respectively. $\zeta=\{\alpha, n, a\}$ is the multiindex.

Since wavefronts of distributions can be localized \cite{Trev82} and a differential
manifold locally resembles Euclidian space near each point, we consider
the case when the space-time manifold ${\bfb M}$ is the
$4$-­dimensional Euclidian (pseudo-Euclidian) space or an open subset of
this space with the Euclidian metric tensor $g^{\mu\nu}$. In order to
derive quantum-field equations, we consider the action of the fields $\Psi$ on the
manifold ${\bfb M}$ \cite{Huang82,Cheng84,Dub92}

$$S(\Psi)=\int_{\bfb M}L(\Psi(x))\,dx $$

\noindent with the Lagrangian

$$L(\Psi(x))=-\frac14 F_{\mu\nu}^{a}(x)F^{\mu\nu a}(x) +
i\bar\psi^{\alpha}(x)\gamma^{\mu}\nabla_{\mu\beta}^{\alpha}\psi^{\beta}(x)$$
$$+\bar\nabla_{\mu n}^l\varphi^{{+}n}(x)\nabla^{\mu
l}_m\varphi^m(x)$$
\begin{equation}
-m(\psi^{\alpha}(x), \bar\psi^{\alpha}(x), \varphi^n(x),
\varphi^{{+}n}(x)), \label{eq2}
\end{equation}

\noindent where

$$F_{\mu\nu}^{a}=\partial_{\mu}A^{a}_{\nu}- \partial_{\nu}A^{a}_{\mu}+
e_{j(a)}C^a_{bc}A^{b}_{\mu}A^{c}_{\nu},$$

$$F^{a\mu\nu}=g^{\mu\rho}g^{\nu\sigma}F_{\rho\sigma}^{a},$$

\noindent are the intensity of the gauge fields,

$$\gamma^{\mu}\gamma^{\nu}+\gamma^{\nu}\gamma^{\mu}=2g^{\mu\nu}$$

\noindent are Dirac matrices,

$$\nabla_{\mu\beta}^{\alpha}=\partial_{\mu}\delta^{\alpha}_{\beta}-
ie_{j(a)}T_{a\beta}^{\alpha}A^{a}_{\mu},$$

$$\nabla_{\mu m}^{l}=\partial_{\mu}\delta^{l}_{m}-
ie_{j(a)}{\tau}_{a m}^{l}A^{a}_{\mu},\qquad \nabla_{m}^{\mu
l}=g^{\mu\nu}\nabla_{\nu m}^{l},$$

$$\bar\nabla_{\mu m}^{l}=\partial_{\mu}\delta^{l}_{m}+
ie_{j(a)}{\tau}_{a m}^{{*}l}A^{a}_{\mu},$$

\noindent $T_a=\|T_{a\beta}^{\alpha}\|$ is the gauge matrix
with the commutation relation $[T_a,T_b]=iC_{ab}^cT_c$ acting on
spinors as $\psi'=\exp(i\sigma^aT_a)\psi$, $\sigma^a$ is an
arbitrary real number. ${\tau}_a=\|{\tau}_{an}^{m}\|$ is the
gauge matrix with the commutation relation
$[{\tau}_a,{\tau}_b]=iC_{ab}^c{\tau}_c$ acting on bosons as
$\varphi'=\exp(i\sigma^a{\tau}_a)\varphi$. It is supposed that the
summation in relation (\ref{eq2}) and in the all following relations
is performed over all repeating indices. $e_{j(a)}$ is the charge
corresponding to the $j$ factor of the direct decomposition of the
gauge group ${\bfb G}=\prod_j{\bfb G}_j$ ($=SU(3)\times
SU(2)\times U(1)$). If the index $a$ of operators $T_a$ and
${\tau}_a$ belongs to the subgroup ${\bfb G}_j$, then in the charge
$e_{j(a)}$ $j(a)=j$. In the Lagrangian (\ref{eq2}) the polynomial
$m(\psi^{\alpha}, \bar\psi^{\alpha}, \varphi^n, \varphi^{{+}n})$
does not contain field derivatives. The polynomial $m$ determines
mass terms and interactions between fields.

For derivation of the quantum-field Schwinger equation it needs to
add the linear term $(Q,\Psi)$ with auxiliary fields
$Q(x)=\{Q^{\zeta}(x)\}=\{Q^{\alpha}_{(\psi)}(x),
Q^{\alpha}_{(\bar\psi)}(x), Q^{n}_{(\varphi)}(x),
Q^{n}_{(\varphi^{+})}(x), Q^{a}_{(A)\mu}(x)\}$ to the action
$S(\Psi)$

$$\bar S(\Psi, Q)= S(\Psi)+(Q,\Psi), $$

\noindent where

$$(Q,\Psi)=\int_{\bfb M}(Q^{\alpha}_{(\psi)}(x)\psi^{\alpha}(x)+
Q^{\alpha}_{(\bar\psi)}(x)\bar\psi^{\alpha}(x)+
Q^{n}_{(\varphi)}(x)\varphi^{n}(x)$$
$$+Q^{n}_{(\varphi^{+})}(x)\varphi^{{+}n}(x)
+Q^{a}_{(A)\mu}(x)A^{a}_{\mu}(x))\,dx.$$

\noindent For fields $\varphi^{n}(x)$, $\varphi^{{+}n}(x)$, and
$A^{a}_{\mu}(x)$ the auxiliary fields $Q^{\zeta}(x)$ are simple
variables and for fields $\psi^{\alpha}(x)$ and
$\bar\psi^{\alpha}(x)$ the auxiliary fields are Grassmanian ones,
respectively. Then, the Schwinger equation is written in the form
\cite{Schw51,Vas98}

\begin{equation}
\left\{\left.\frac{\vec\delta S(\Psi)}{\delta\Psi^{\zeta}(x)}
\right|_{\Psi^{\zeta}(x)=\kappa\hbar\vec\delta/\delta iQ^{\zeta}(x)}
+Q^{\zeta}(x)\right\}G(Q)=0, \label{eq3}
\end{equation}

\noindent where $\vec\delta/\delta\Psi$ is the functional derivative
on the left, $\kappa=1$ for bosons and $\kappa=-1$ for fermions,
$G(Q)$ is the generating functional. The formal solution of the
Schwinger equation (\ref{eq3}) is the functional integral

$$G(Q)=\int\exp\left[\frac{i}{\hbar}\left(S(\Psi) +(Q,\Psi)\right)\right]\,D\Psi.$$

\noindent We consider densities of the quantum fields
$\psi^{\alpha}(x)$, $\bar\psi^{\alpha}(x)$, $\varphi^{n}(x)$,
$\varphi^{{+}n}(x)$, and $A^{a}_{\mu}(x)$

$$W^{\zeta}(x)=\left.\frac{\vec\delta G(Q)}{\delta Q^{\zeta}(x)}
\right|_{Q\rightarrow 0}.$$

\noindent Taking into account the form of the Lagrangian
(\ref{eq2}), from the Schwinger equation (\ref{eq3}) we can obtain
differential equations for the densities
$W(x)=\{W^{\zeta}(x)\}=\{W^{\alpha}_{(\psi)}(x),
W^{\alpha}_{(\bar\psi)}(x), W^{n}_{(\varphi)}(x),
W^{n}_{(\varphi^{+})}(x), W^{a}_{(A)\mu}(x)\}$, which can be written
in the form

$$W^{\alpha}_{(\bar\psi)}(x)\left(\gamma^{\mu}\overleftarrow{\partial}_{\mu}
-\frac{m^{(\alpha)}c}{\hbar}\right)=
B^{\alpha}_{(\bar\psi)}(R^{(s)}(x))$$

$$\left(\gamma^{\mu}\vec{\partial}_{\mu}
+\frac{m^{(\alpha)}c}{\hbar}\right)W^{\alpha}_{(\psi)}(x)=
B^{\alpha}_{(\psi)}(R^{(s)}(x))$$

$$\left(\Box -\frac{m^{(n)2}c^2}{\hbar^2}\right)W^{n}_{(\varphi)}(x)=
B^{n}_{(\varphi)}(R^{(s)}(x))$$

$$\left(\Box -\frac{m^{(n)2}c^2}{\hbar^2}\right)
W^{n}_{(\varphi^{+})}(x)= B^{n}_{(\varphi^{+})}(R^{(s)}(x))$$

\begin{equation}
\Box W^{a}_{(A)\mu}(x)= B^{a}_{(A)\mu}(R^{(s)}(x)), \label{eq4}
\end{equation}

\noindent where

$$\Box (\cdot)= g^{\mu\nu}\frac{\partial^2{(\cdot)}}{\partial
x^{\mu}x^{\nu}}$$

\noindent is the d'Alembert operator; $m^{(\alpha)}$ and $m^{(n)}$
are masses of fermions and bosons, respectively;
$B^{\zeta}=\{B^{\alpha}_{(\psi)}, B^{\alpha}_{(\bar\psi)},
B^{n}_{(\varphi)}, B^{n}_{(\varphi^{+})}, B^{a}_{(A)\mu}\}$ are
polynomials of higher order Green's functions

$$R^{(s)}(x)=\left.\frac{\vec\delta^s G(Q)}{\underbrace{\delta Q^{\zeta}(x)\ldots
\delta Q^{\eta}(x)}_s} \right|_{Q\rightarrow 0}, \qquad (s>1).$$

Equations (\ref{eq4}) can be written in the form

\begin{equation}
{\bfb N}_{\zeta}^{\eta}W^{\zeta}(x)=B^{\eta}(R^{(s)}(x)),
\label{eq5}
\end{equation}

\noindent where ${\bfb N}_{\zeta}^{\eta}=
\sum_{q=0}^ka_{q\zeta}^{\eta}\partial_x^q$ is the matrix differential operator.
Solutions of the Schwinger equation (\ref{eq3}) determined the generating functional
$G(Q)$ and solutions of equations (\ref{eq4}), (\ref{eq5}) can be found in the
approximate form by the diagram technique~\cite{Vas98,Lut07,Lut09}. We will not
find solutions of the Schwinger equation. Our aim is to analyse singularities of solutions.
For our purposes it is sufficient to note
that, in the common case, solutions of equations (\ref{eq4}) and (\ref{eq5}) are
distributions. The question is: which distributions of the densities $W^{\zeta}(x)$
can be multiplied and, therefore, form a commutative algebra? We suppose that
the densities $W^{\zeta}(x)$ can be expressed in the oscillatory--integral
form~\cite{Trev82,Hor85,Hor4}

\begin{equation}
W^{\zeta}(x)=\int_{T^{*}{\bfb M}}F^{\zeta}(x,\chi)
\exp{[i\sigma^{\zeta}(x,\chi)]}\,d\chi, \label{eq6}
\end{equation}

\noindent where $T^{*}{\bfb M}$ is the cotangent bundle over the
space-time manifold ${\bfb M}$, $\sigma^{\zeta}(x,\chi)$ is the
phase function, $F^{\zeta}(x,\chi)$ is the amplitude, $\chi$ is the
covector, and $(x,\chi)\in T^{*}{\bfb M}$. It needs to note that
relation (\ref{eq6}) defines Lagrangian distributions, which form
the subset of the space of all distributions $D'(\bfb M)$. Any
Lagrangian distribution can be represented locally by oscillatory
integrals~\cite{Hor4}. Conversely, any oscillatory integral is a
Lagrangian distribution. We consider the case of the real linear
phase function of $\chi$

\begin{equation}
\sigma^{\zeta}(x,\chi)=k^{\zeta}{\chi}_{\nu}x_{\nu},
\label{eq7}
\end{equation}

\noindent where $k^{\zeta}$ is a coefficient. Our consideration of
the case of Lagrangian distributions is
motivated by the statement that, if the multiplication cannot be
defined on the Lagrangian distribution subset, then this operation
cannot be defined on the space $D'({\bfb M})$.

The wavefront set $\WF(u)$ of a distribution $u$ can be defined as
\cite{Hor71,Hor85,Trev82}

$$\WF(u)=\{(x,\xi)\in T^{*}{\bfb M}|\; \xi\in\Gamma_x(u)\},$$

\noindent where the singular cone $\Gamma_x(u)$ is the complement of
all directions $\xi$ such that the Fourier transform of $u$,
localized at $x$, is sufficiently regular when restricted to a
conical neighborhood of $\xi$. The wavefront of the distributions
$\WF(W^{\zeta})$ characterizes the singularities of solutions and
is determined by the wavefront of $R^{(s)}(x)$ and by the
characteristics of the matrix operator ${\bfb N}_{\zeta}^{\eta}$
\cite{Hor85,Trev82}

$$\WF(W^{\zeta})\subset\Char({\bfb N}_{\zeta}^{\eta})
\bigcup\WF(B^{\eta}(R^{(s)})), $$

\noindent where the characteristics $\Char({\bfb N}_{\zeta}^{\eta})$
is the set $\{(x,\xi)\in T^{*}{\bfb M}\setminus 0\}$ defined by
linear algebraic equations of highest power orders in the unknown
Fourier transforms $\tilde W^{\zeta}(\xi)$

$$\sum_{\zeta, q=\max k}a_{q\zeta}^{\eta}(-i\xi)^q\tilde W^{\zeta}(\xi)=0.$$

\noindent The covector $\xi\in T^{*}(x)$ lies in the cotangent cone
$\Gamma_x$ at the point $x$. Taking into account equations
(\ref{eq4}), we can find that

$$\left.\Char({\bfb N}_{\zeta}^{\eta})\right|_x=
g^{\mu\nu}\xi_{\mu}\xi_{\nu}, $$

\noindent consequently, singularities of solutions are located on
this cone and partially can be formed by wavefronts of higher order
Green's functions $R^{(s)}$.

Starting from the proposition that properties of the space-time
manifold ${\bfb M}$ are defined by the quantum fields
$\Psi^{\zeta}(x)$ and, consequently, by the commutative algebra
${\bfb A}$ of distributions $W^{\zeta}(x)$, in the next section we
find that this statement results in essential restrictions imposed
on the space-time manifold ${\bfb M}$.

\section{ALGEBRA OF DISTRIBUTIONS OF QUANTUM-FIELD DENSITIES}

In order to determine points of the space-time manifold ${\bfb M}$
by means of the densities $W^{\zeta}(x)$, it is necessary to define
multiplication on densities, to construct the commutative algebra
${\bfb A}$ of distributions of densities, and to find maximal ideals
of this algebra. According to Ref. \cite{Hor85,Trev82}, multiplication on
distributions $u, v\in D'({\bfb M})$ with wavefronts
$\WF(u)=\{(x,\xi)\}$ and $\WF(v)=\{(x,\eta)\}$ is determined, if and
only if

\begin{equation}
\WF(u)\bigcap \WF'(v)=\emptyset, \label{eq8}
\end{equation}

\noindent where $\WF'(v)$ is the image of $\WF(v)$ in the
transformation $(x,\eta)\mapsto (x,-\eta)$ in the cone subset
$\Gamma$ of the cotangent bundle $T^{*}{\bfb M}$. The wavefront of
the product is defined as

$$\WF(uv)\subset\left\{(x,\xi+\eta)|\; (x,\xi)\in\WF(u)\right.$$
\begin{equation}
\left.\mbox{ or } \xi=0, (x,\eta)\in\WF(v) \mbox{ or } \eta=0; \;
\xi+\eta\neq 0\right\}. \label{eq9}
\end{equation}

\noindent Taking into account the linear form of the phase function
(\ref{eq7}), from relations (\ref{eq8}) and (\ref{eq9}) we obtain
restrictions on the densities $W^{\zeta}(x)$. For this purpose, we
consider wavefronts of densities for the cases of the space-time
manifold ${\bfb M}$, when the structure group of the cotangent
bundle $T^{*}{\bfb M}$ is the Lie group $SO(4-p,p)$ with $p=0$,
$p=1$, and $p>1$.

The fields $\Psi$ (\ref{eq1}) are observed relative to inertial
frames of reference in the space-time manifold ${\bfb M}$.
Rectilinear motion transforms the fields $\Psi$ and, consequently, their densities $W^{\zeta}(x)$ and
wavefronts. This transformation can be considered as a
diffeomorphism of ${\bfb M}$ and is described by the arcwise
connected part of the $SO(4-p,p)$ group. More precisely, if
$f:\Omega\rightarrow{\Omega}'$ is the diffeomorphism of open subsets
$\Omega,{\Omega}'\subset{\bfb M}$, $f_{+}:T^{*}\Omega\rightarrow
T^{*}{\Omega}'$ is the diffeomorphism induced by $f$, and
$f_{*}:D'(\Omega)\rightarrow D'({\Omega}')$ is the isomorphism, then
for the distribution $u\in D'(\Omega)$~\cite{Hor85,Trev82}

\begin{equation}
\WF(f_{*}u)=f_{+}\WF(u). \label{eq10}
\end{equation}

\noindent Thus, rectilinear motion results in the transformation
$f_{*}$ of the field densities $W^{\zeta}(x)$ induced by the arcwise
connected part of the structure group and the transformation $f_{+}$
of density wavefronts (\ref{eq10}). If, as a result of these
transformations, the covector $\xi\in\WF(W^{\zeta}) \subset
T^{*}{\bfb M}$ changes its orientation $\xi\mapsto -\xi$, then the
multiplication (\ref{eq9}) on the density $W_1^{\zeta}(x)$ with the
covector $\xi$ and the density $W_2^{\zeta}(x)$ with the covector
$-\xi$ is impossible. We consider transformations of density
wavefronts induced by the arcwise connected part of the structure
group (\ref{eq10}) and by discrete symmetry transformations -- the
time reversal and the space reflection.

\subsection{Time reversal}

We assume that the co-ordinate variables of the space-time manifold
${\bfb M}$ can be divided by time $t$ and space $r$ variables,
$x=\{t_1,\ldots,t_p,r_1,\ldots\}$ $(p=0,1,2,\ldots)$. The structure
group of the space-time manifold is $SO(4-p,p)$. We consider the
time reversal $T$ on the manifold ${\bfb M}$ and the possibility of
embedding of the time reversal into the arcwise connected part of
the group $SO(4-p,p)$.

\vspace{10pt}

\begin{figure}
\begin{center}
\includegraphics*[scale=0.6]{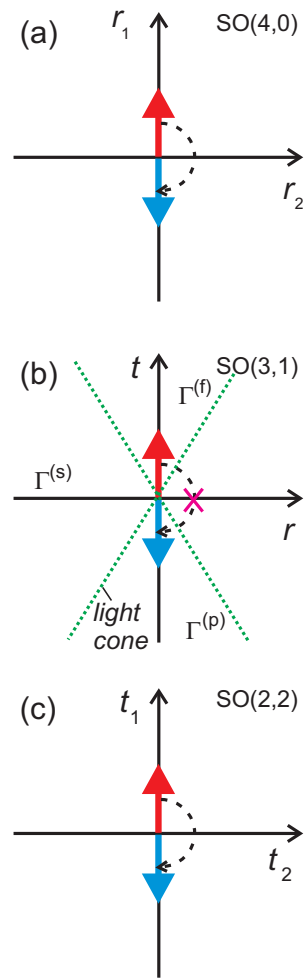}
\end{center}
\caption{Transformation of the covector $\xi\mapsto -\xi$ at the
time reversal of the space-time manifold ${\bfb M}$ with the
structure groups (a) $SO(4,0)$, (b) $SO(3,1)$, and (c) $SO(2,2)$.
The image of the transformation $\xi\mapsto -\xi$ can be reached by
means of a transformation of the arcwise connected part of groups
$SO(4,0)$ and $SO(2,2)$ and is not attainable for the case of the
arcwise connected part of the $SO(3,1)$ group.} \label{Fig1}
\end{figure}

\noindent {\bf Euclidian space-time manifold with the structure
group $SO(4,0)$}. Let us consider the case of the time reversal on
the Euclidian space-time manifold ${\bfb M}$ with the structure
group $SO(4,0)$ of the cotangent bundle $T^{*}{\bfb M}$. The
signature of the space-time metric is $({-}{-}{-}{-})$. In this
case, $p=0$ and the time variable $t$ is identical to the space one
(for example, $r_1$). The image of the inversion $\xi\mapsto -\xi$
in relation (\ref{eq10}) on wavefronts of densities can be reached
by the transformation induced by the arcwise connected part of the
group $SO(4,0)$ (Fig. \ref{Fig1}a). Thus, we always can find
densities with covectors $\xi$ and $\eta$ in relation (\ref{eq9})
such that $\xi+\eta=0$. Consequently, multiplication on
distributions in the Euclidian space-time manifold with the
signature of the space-time metric $({-}{-}{-}{-})$ is impossible.
One can only say about a partial multiplication operation on a
subset of the distribution space. The analogous consideration can be
carried out for manifold ${\bfb M}$ with the structure group
$SO(0,4)$ and with the signature of the space-time metric
$({+}{+}{+}{+})$.

\vspace{10pt}

\noindent {\bf Pseudo-Euclidian space-time manifold with the
structure group $SO(3,1)$}. The signature of the space-time metric
is $({+}{-}{-}{-})$. In accordance with relation
$g^{\mu\nu}\xi_{\mu}\xi_{\nu}=0$ defined boundaries of a singular
cone, the space-time manifold is separated by three regions: the
future light cone $\Gamma^{(f)}$, the past light cone
$\Gamma^{(p)}$, and the space-like region $\Gamma^{(s)}$ (Fig.
\ref{Fig1}b). The multiplication of distributions with singularities
in the region $\Gamma^{(s)}$ is impossible because of the existence
of the inversion $\xi\mapsto -\xi$ reached by the arcwise connected
part of the group $SO(3,1)$. Therefore, singularities can exist only
in the cone $\Gamma^{(f)}$ or in the cone $\Gamma^{(p)}$. The time
reversal $T$ inverts the covector $\xi\in\WF(W^{\zeta}) \subset
T^{*}{\bfb M}$ from the future light cone $\Gamma^{(f)}$ to the past
light cone $\Gamma^{(p)}$ (Fig. \ref{Fig1}b). The Lorentz
transformation $SO(3,1)$ acts on future and past light cones
separately. Consequently, the arcwise connected part of the group
$SO(3,1)$ does not contain the time inversion. So, if we exclude
field density distributions $W^{\zeta}(x)$ with singularities in the
past light cone $\Gamma^{(p)}$ and consider only density
distributions with singularities in the future light cone
$\Gamma^{(f)}$, then for these distributions the multiplication can
be defined. Correspondingly, densities of states $T\Psi^{\zeta}(x)$
are forbidden. In this case, we have the one-way direction of time
and there is not the symmetry of time on density distributions. The
arrow of time is pointing towards the future. According to
(\ref{eq9}), the wavefront of the product of quantum-field densities
in the future light cone $\Gamma^{(f)}$ is given by

$$\WF(W^{\zeta_1}\cdot\ldots\cdot W^{\zeta_n})\subset
\{(x,\sum_i^n\xi^{\zeta_i})| \; \xi^{\zeta_i}\in \Gamma^{(f)}, \;
\sum_i^n \xi^{\zeta_i}\neq 0\},$$

\noindent where $\xi^{\zeta_i}=\beta^{\zeta_i}\chi^{(i)}$ is the
covector associated with the density $W^{\zeta_i}$, $\beta^{\zeta}$ is a function
of field invariants, such as chirality, charge signs, charge parity,
etc. Taking into
account that for the Dirac adjoint spinor density
$W^{\alpha}_{(\bar\psi)}(x)$ and for the density
$W^{n}_{(\varphi^{+})}(x)$ the exponent term in the integral in
relation (\ref{eq6}) is transformed as $i\sigma^{\zeta}(x,\chi)
\rightarrow -i\sigma^{\zeta}(x,\chi)$, we find that in the cone
$\Gamma^{(f)}$

$$\beta_{(\psi)}>0 \qquad \beta_{(\bar\psi)}<0$$
\begin{equation}
\beta_{(\varphi)} >0 \qquad \beta_{(\varphi^{{+}})}<0. \label{eq11}
\end{equation}

\vspace{10pt}

\noindent {\bf Pseudo-Euclidian space-time manifold with the
structure group $SO(4-p,p)$ ($p>1$)}. The signatures of the
space-time metric are $({+}{+}{-}{-})$ ($p=2)$ and $({+}{+}{+}{-})$
($p=3)$. If the time dimension is equal to 2 or higher ($p>1$), then
the image of the time inversion of densities in the time plane
$(t_1,\ldots , t_p)$ is attained by means of a transformation of the
arcwise connected part of the group $SO(4-p,p)$ (Fig. \ref{Fig1}c).
In this case, one can find densities with covectors $\xi$ and $\eta$
in relation (\ref{eq9}) such that $\xi+\eta=0$. Multiplication on
distributions in this space-time manifold is impossible.

\subsection{Space reflection and charge conjugation in the fermion sector}

Another restriction to define multiplication on the density
distribution algebra is caused by the chirality of fermions and by
the charge conjugation. Since, according to the above-mentioned
subsection the pseudo-Euclidian space-time manifold with the
Lorentz group $SO(3,1)$ and the arrow of time are necessary
conditions for definition of multiplication on density
distributions, we consider fermion distributions with singularities
in the future light cone $\Gamma^{(f)}$. Right­-handed and
left­-handed states of the Dirac fields $\psi^{\alpha}(x)$ and
$\bar\psi^{\alpha}(x)$ are defined by projective operators $(1\pm
\gamma^5)/2$ acting on a spinor \cite{Huang82,Cheng84,Dav}

$$\psi^{\alpha}_R(x)= \frac{1+ \gamma^5}{2}\psi^{\alpha}(x)$$

$$\psi^{\alpha}_L(x)= \frac{1- \gamma^5}{2}\psi^{\alpha}(x).$$

\noindent The space reflection $P$ converts the right­-handed spinor into
the left­-handed one, and vice versa

$$P\psi^{\alpha}_R(x)=iL_p\psi^{\alpha}_L(x)$$
$$P\psi^{\alpha}_L(x)=iL_p\psi^{\alpha}_R(x),$$

\noindent where in the Weyl (chiral) basis

$$L_p=\left( \begin{array}{cc}
0&I\\I&0 \end{array}\right)$$

\noindent and $I$ is the identity 2-matrix. $\psi^{\alpha}_R(x)$
and $\psi^{\alpha}_L(x)$ are eigenvectors of the operator $\gamma^5$
with the chirality $\lambda=1$ (the right­-handed state) and the
chirality $\lambda=-1$ (the left­-handed state), respectively.
Double space reflection $P^2$ can be regarded as $360^{\circ}$
rotation. It transforms spinors as

$$P^2\psi^{\alpha}_{\rho}(x)=-\psi^{\alpha}_{\rho}(x),$$

\noindent where $\rho=\{R,L\}$.

For fulfilment of relations (\ref{eq11}) the coefficients
$\beta_{(\psi)}$ and $\beta_{(\bar\psi)}$ must contain a charge
factor $\kappa_c$ such that $\beta(\psi)=1$ and
$\beta(\bar\psi)=-1$. We consider two cases, (1) $\beta
=\lambda\kappa_c$ with the chirality $\lambda$ and (2) $\beta
=\kappa_c$ without chirality. In order to fulfil multiplication on
densities $W^{\alpha}_{(\psi_R)}(x)$, $W^{\alpha}_{(\psi_L)}(x)$,
$W^{\alpha}_{(\bar\psi_R)}(x)$, $W^{\alpha}_{(\bar\psi_L)}(x)$ in
the first case, for right­-handed and left­-handed states of the
fermion fields we should get

$$\kappa_c(\psi^{\alpha}_R(x))=1, \qquad
\kappa_c(\psi^{\alpha}_L(x))=-1, $$
$$\kappa_c(\bar\psi^{\alpha}_R(x))=-1, \qquad
\kappa_c(\bar\psi^{\alpha}_L(x))=1,$$

$$\lambda(\psi^{\alpha}_R(x))=1, \qquad
\lambda(\psi^{\alpha}_L(x))=-1,$$
$$\lambda(\bar\psi^{\alpha}_R(x))=1, \qquad
\lambda(\bar\psi^{\alpha}_L(x))=-1. $$

\noindent The charge conjugation $C$ transforms $\kappa_c$:
$C\kappa_c(\psi^{\alpha}_R(x))=
\kappa_c(\bar\psi^{\alpha}_R(x))=-1$,
$C\kappa_c(\bar\psi^{\alpha}_R(x))= \kappa_c(\psi^{\alpha}_R(x))=1$,
$C\kappa_c(\psi^{\alpha}_L(x))= \kappa_c(\bar\psi^{\alpha}_L(x))=1$,
and $C\kappa_c(\bar\psi^{\alpha}_L(x))=
\kappa_c(\psi^{\alpha}_L(x))=-1$. The commutative algebra ${\bfb A}$
of distributions contains densities $W^{\zeta}(x)$ of states
$\psi^{\alpha}_R(x)$, $\psi^{\alpha}_L(x)$,
$\bar\psi^{\alpha}_R(x)$, $\bar\psi^{\alpha}_L(x)$,
$CP\psi^{\alpha}_R(x)$, $CP\psi^{\alpha}_L(x)$,
$CP\bar\psi^{\alpha}_R(x)$, $CP\bar\psi^{\alpha}_L(x)$, their sums
and products. Densities $W^{\zeta}(x)$ of states
$P\psi^{\alpha}_R(x)$, $P\psi^{\alpha}_L(x)$,
$P\bar\psi^{\alpha}_R(x)$, $P\bar\psi^{\alpha}_L(x)$,
$C\psi^{\alpha}_R(x)$, $C\psi^{\alpha}_L(x)$,
$C\bar\psi^{\alpha}_R(x)$, $C\bar\psi^{\alpha}_L(x)$ are forbidden
and are not contained in the algebra ${\bfb A}$. This version of the
theoretical model can explain the chirality violation.

In the second case, the fulfilment of the relation $\beta =\kappa_c$
with $\kappa_c(\psi^{\alpha}_R(x))=\kappa_c(\psi^{\alpha}_L(x))=1$
and $\kappa_c(\bar\psi^{\alpha}_R(x))
=\kappa_c(\bar\psi^{\alpha}_L(x))=-1$ results in forbidden densities
$W^{\zeta}(x)$ of states $C\psi^{\alpha}_R(x)$,
$C\psi^{\alpha}_L(x)$, $C\bar\psi^{\alpha}_R(x)$, and
$C\bar\psi^{\alpha}_L(x)$. Wavefronts of these densities are in the
past light cone $\Gamma^{(p)}$: $\WF(W^{\alpha}_{(C\psi_R)})$,
$\WF(W^{\alpha}_{(C\psi_L)})$, $\WF(W^{\alpha}_{(C\bar\psi_R)})$,
and $\WF(W^{\alpha}_{(C\bar\psi_L)})\in\Gamma^{(p)}$. At the same
time, the commutative algebra ${\bfb A}$ of distributions contains
densities $W^{\zeta}(x)$ of states $\psi^{\alpha}_R(x)$,
$\psi^{\alpha}_L(x)$, $\bar\psi^{\alpha}_R(x)$,
$\bar\psi^{\alpha}_L(x)$, $P\psi^{\alpha}_R(x)$,
$P\psi^{\alpha}_L(x)$, $P\bar\psi^{\alpha}_R(x)$, and
$P\bar\psi^{\alpha}_L(x)$. The chirality is not violated. In the
experiment the chirality violation of fermions is observed and,
therefore, this case of the theoretical model must be ignored.

\subsection{Charge conjugation in the boson sector}

We consider the common case of the boson (Higgs) sector containing
the quantum fields $\varphi^n(x)$ and $\varphi^{{+}n}(x)$. We assume
that $\varphi^n(x)\neq\varphi^{{+}n}(x)$. By analogy with the
fermion case, the covector of singularity of the density of
$\varphi^n(x)$ is $\beta_{(\varphi)}\chi$ and the analogous covector
of the density of $\varphi^{{+}n}(x)$ is
$-\beta_{(\varphi^{+})} \chi$, respectively. In order to
fulfil relations (\ref{eq11}) and the requirement that
$\beta_{(\varphi)}\chi$, $-\beta_{(\varphi^{+})} \chi\in
\Gamma^{(f)}$, we must write the coefficient $\beta$ in the form
$\beta_{(\varphi)} =\kappa_c(\varphi^n(x))$ and
$\beta_{(\varphi^{+})} =\kappa_c(\varphi^{{+}n}(x))$. The charge
conjugation $C$ changes the sign of the factor $\kappa_c$:
$C\kappa_c(\varphi^n(x))= \kappa_c(\varphi^{{+}n}(x))=-1$ and
$C\kappa_c(\varphi^{{+}n}(x))= \kappa_c(\varphi^n(x))=1$. Thus,
densities of $\varphi^n(x)$ and $\varphi^{{+}n}(x)$ can be
multiplied and are included in the algebra ${\bfb A}$. On the
contrary, wavefronts of densities of states $C\varphi^n(x)$ and
$C\varphi^{{+}n}(x)$ are in the past light cone $\Gamma^{(p)}$ and
must be excluded. This leads to the charge conjugation symmetry
violation in the boson sector.

It needs to note that in the modified theoretical models of the
Higgs boson sector
\cite{Gun03,Bran12,Hab85,Baer06,Shu13,Cir16,Car16,Engl17,Good17}
extending the BEH model \cite{Engl64,Higgs64,Higgs64a} some
particles in the Higgs sector have negative charge parities and are
charged. In this case, the above-mentioned $C$-violation on density
distributions in the Higgs sector can explain the observed
matter-antimatter imbalance.

Thus, as a result of this section, we define the multiplication on
distributions of the quantum-field densities and construct the
commutative density distribution algebra ${\bfb A}$. The algebra
${\bfb A}$ are generated by distributions with singularities in the
future light cone $\Gamma^{(f)}$. The asymmetry of time
($T$-violation), the chiral asymmetry ($P$-violation) and the charge
($C$) conjugation symmetry violation are caused by singularities of
density distributions and these space-time manifold properties are
local. Spectrum of the density distribution algebra ${\bfb A}$ and
the scheme $({\bfb M},{\bfb A}_{{\bfb M}})$ are considered in the
next section.

\section{IDEALS AND SPECTRUM OF THE DENSITY DISTRIBUTION ALGEBRA.
SCHEME $({\bfb M},{\bfb A}_{{\bfb M}})$}

For definition of the scheme $({\bfb M},{\bfb A}_{{\bfb M}})$
contained the spectrum of the commutative density distribution
algebra ${\bfb A}$ isomorphic to the space-time manifold ${\bfb M}$,
it needs to carry out localization and to determine the sheaf of
structure algebras and the spectrum of the algebra ${\bfb A}$. To
this end, we define prime and maximal ideals of this algebra. The
algebra ${\bfb A}$ consists of density distributions $W^{\zeta}(x)$
with singularities in the closed future light cone subset,
$\WF(W^{\zeta})\subset\Gamma^{(f)} \subset T^{*}{\bfb M}$. The
complement of the future light cone subset $\Gamma^{(f)}$ is the
open cone subset $\bar\Gamma^{(f)}$. In the cone subset
$\bar\Gamma^{(f)}$ the densities $W^{\zeta}(x)$ are
$C^{\infty}$-smooth functions. We define the maximal ideal at the
point $x_0$ as the set of distributions equal to zero at the point
$x_0$ in the cone subset $\bar\Gamma^{(f)}$

$$m_{x_0}= \{W^{\zeta}(x)| \; \lim_{x\rightarrow x_0} W^{\zeta}(x)=0 \quad \mbox{in}
\quad \bar\Gamma^{(f)}\}.$$

\noindent $p$ is called a prime ideal if for all distributions
$W_1(x)$ and $W_2(x)\in{\bfb A}$ with $W_1(x)W_2(x)\in p$ we have
$W_1(x)\in p$ or $W_2(x)\in p$ \cite{Cox98,Lang02}. Every maximal
ideal $m_x$ is prime.

In the process of localization of the algebra ${\bfb A}$ we find a
local algebra contained only information about the behavior of
density distributions $W^{\zeta}(x)$ near the point $x$ of the
space-time manifold ${\bfb M}$. We consider the case of maximal
ideals. Then, the local algebra ${\bfb A}_x$ is defined as the
commutative algebra consisting of fractions of density distributions
$W_1(x)$ and $W_2(x)$

\begin{equation}
{\bfb A}_x=\left\{\left.\frac{W_1(x)}{W_2(x)}\right| \;
W_1(x),W_2(x)\in{\bfb A},\; W_2(x)\notin m_x\right\}, \label{eq12}
\end{equation}

\noindent where $m_x$ is the maximal ideal at the point $x$. The
fraction $W_1(x)/W_2(x)$ is the equivalence class defined as

$$\frac{W_1(x)}{W_2(x)}=\frac{W'_1(x)}{W'_2(x)},$$

\noindent if there exists the distribution $V(x)\in{\bfb A}$,
$V(x)\notin m_x$ such that

$$V(x)(W_1(x)W'_2(x)-W_2(x)W'_1(x))=0. $$

\noindent Operations on the local algebra ${\bfb A}_x$ (\ref{eq12})
look identical to those of elementary algebra

$$\frac{W_1(x)}{W_2(x)}+\frac{W'_1(x)}{W'_2(x)}=
\frac{W_1(x)W'_2(x)+W'_1(x)W_2(x)}{W_2(x)W'_2(x)}$$

\noindent and

$$\frac{W_1(x)}{W_2(x)}\cdot\frac{W'_1(x)}{W'_2(x)}=
\frac{W_1(x)W'_1(x)}{W_2(x)W'_2(x)}.$$

\noindent Algebras ${\bfb A}_{U_i}$ on open sets  $U_i\subset {\bfb
M}$

$${\bfb A}_{U_i}=\bigcap_{x\in U_i}{\bfb A}_x $$

\noindent determine the structure sheaf \cite{Mum99} ${\bfb
A}_{{\bfb M}}=\{{\bfb A}_{U_i}\}$ on the space-time manifold ${\bfb
M}$. The inverse limit of the structure sheaf ${\bfb A}_{{\bfb M}}$
coincides with the algebra ${\bfb A}$

$${\bfb A}=\lim_{\overleftarrow{U_i\in{\bfb M}}}{\bfb A}_{U_i},$$

\noindent where $\{U_i\}$ is the open covering of ${\bfb M}$.

The spectrum of the algebra ${\bfb A}$, denoted by $\Spec({\bfb
A})$, is the set of all prime ideals of ${\bfb A}$, equipped with 
the Zariski topology \cite{Mum99,EGA,Eis98}. The prime ideals
correspond to irreducible subvarieties of the space $\Spec({\bfb
A})$. Maximal ideals of the algebra ${\bfb A}$ correspond to points.

The structure sheaf and the spectrum of the algebra ${\bfb A}$ are
used in definition of schemes~\cite{Mum99,EGA,Eis98}. In our case,
the scheme over the algebra ${\bfb A}$ is the pair $({\bfb M},{\bfb
A}_{{\bfb M}})$ such that there exists an open covering $\{U_i\}$ of
${\bfb M}$ for which each pair $(U_i,{\bfb A}_{U_i})$ is isomorphic
to $(V_i,{\bfb O}_{V_i})$, where $\{V_i\}$ is the open covering of
$\Spec({\bfb A})$ and ${\bfb O}_{V_i}$ is the restriction of the
structure sheaf $\bfb O_{\Spec({\bfb A})}$ to each $V_i$. As a
result, one can say that the local isomorphism ${\bfb
M}\cong\Spec({\bfb A})$ imposed by the theory of schemes and by
restrictions on multiplication on the quantum-field-density
distributions in the algebra ${\bfb A}$ lead to the dependence of
the space-time properties on the matter. The arrow of time, the
chirality violation of spinor fields, and the charge conjugation
symmetry violation in the boson sector are consequences of this
dependence.

\section{CONCLUSION}

In summary, in this paper we describe the dependence between quantum
fields and space-time properties in the framework of the
scheme theory. Contrary to algebras of smooth
functions, densities of quantum fields, which can be found from the
Schwinger equation, are distributions and, in the common case, do
not form an algebra. In order to determine the commutative algebra
${\bfb A}$ of distributions of quantum-field densities, ideals and
its spectrum, it is necessary to define multiplication on densities
and to eliminate those densities, which cannot be multiplied. This
leads to essential restrictions imposed on densities forming the
algebra ${\bfb A}$. Taking into account that in the framework of the
scheme theory the space-time manifold ${\bfb M}$ is locally
isomorphic to the spectrum of the algebra ${\bfb A}$, ${\bfb
M}\cong\Spec({\bfb A})$, the restrictions caused by the possibility
to define multiplication on the density distributions result in the
following properties of the space-time manifold ${\bfb M}$.

\noindent (1) The only possible case, when the commutative algebra
${\bfb A}$ of distributions of quantum-field densities exist, is the
case, when the quantum fields are in the space-time manifold ${\bfb
M}$ with the structure group $SO(3,1)$ (Lorentz group). On account
of the local isomorphism ${\bfb M}\cong\Spec({\bfb A})$, the quantum
fields exist only in the space-time manifold with the
one-dimensional time.

\noindent (2) We must exclude field density distributions with
singularities in the past light cone $\Gamma^{(p)}$. The algebra
${\bfb A}$ consists of the density distributions $W^{\zeta}(x)$ with
wavefronts in the closed future light cone subset,
$\WF(W^{\zeta}(x))\subset\Gamma^{(f)} \subset T^{*}{\bfb M}$. In
this case, we have the one-way direction of time and there is not
the symmetry of time on the density distributions. The arrow of time
is pointing towards the future.

\noindent (3) The restrictions caused by multiplication on the
density distributions can explain the chirality violation of spinor
fields. The densities of right­-handed and left­-handed fermion
states $P\psi^{\alpha}_R(x)$, $P\psi^{\alpha}_L(x)$,
$P\bar\psi^{\alpha}_R(x)$, $P\bar\psi^{\alpha}_L(x)$,
$C\psi^{\alpha}_R(x)$, $C\psi^{\alpha}_L(x)$,
$C\bar\psi^{\alpha}_R(x)$, $C\bar\psi^{\alpha}_L(x)$, where $P$ is
the space reflection and $C$ is the charge conjugation, are
forbidden and are not contained in the algebra ${\bfb A}$.

\noindent (4) For bosons (for example, in the Higgs sector) the
densities of states $C\varphi^n(x)$ and $C\varphi^{{+}n}(x)$ must be
excluded from the algebra ${\bfb A}$. This leads to the charge
conjugation symmetry violation and can explain the observed
matter-antimatter imbalance.

\end{document}